\documentclass[twocolumn,prc,preprintnumbers,nofootinbib,showpacs]{revtex4}
\usepackage{graphicx}
\usepackage{bm}
\usepackage{amssymb,amsmath}

\usepackage{array}

\def\beqn{\begin{eqnarray}}
\def\eeqn{\end{eqnarray}}
\def\barr{\begin{array}}
\def\earr{\end{array}}
\def\btab{\begin{tabular}}
\def\etab{\end{tabular}}
\def\bite{\begin{itemize}}
\def\eite{\end{itemize}}
\def\bcen{\begin{center}}
\def\ecen{\end{center}}

\def\eq{\begin{equation}}
\def\ee{\end{equation}}
\def\nn{\nonumber}

\def\pgdagger{P\hspace{-0.24cm}/}

\def\keldagger{k\hspace{-0.2cm}/}

\def\q2dagger{q_2\hspace{-0.35cm}/\;}

\begin{document}


\title{Dispersion $\gamma Z$-box correction to the weak charge of the proton}

\author{M. Gorchtein} 
\email{mgorshte@indiana.edu}
\author{C. J. Horowitz} 
\email{horowit@indiana.edu}
\affiliation{Nuclear Theory Center and Department of Physics, 
Indiana University, Bloomington, IN 47408, USA} 

\date{\today}

\begin{abstract}
We consider elastic scattering of electrons off a proton target. The parity 
violating (PV) asymmetry arises at leading order in $\alpha$ due to 
interference of $\gamma$ and $Z$ exchange. The radiative corrections to this 
leading mechanism were calculated in the literature and included in 
experimental analyses, except for $\gamma Z$ box and 
cross-box contributions. We present here a dispersion calculation of these 
corrections in forward kinematics. We demonstrate that at the GeV energies of current PV experiments, such corrections are not suppressed by the small vector weak charge of the electron, as occurs in the atomic PV.   Our results suggest that the current theoretical uncertainty in the analysis of the QWEAK experiment might be substantially underestimated, and more accurate account of the dispersion corrections are needed in order to interprete the PV data. 
\end{abstract}


\maketitle

\begin{figure}[h]
{\includegraphics[height=1.2cm]{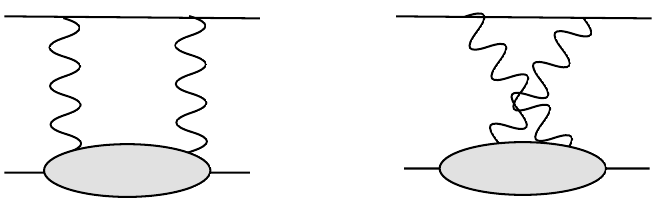}}
\caption{Direct and crossed box diagrams}
\label{fig:tbe}
\end{figure}

\begin{figure}[h]
{\includegraphics[height=1.2cm]{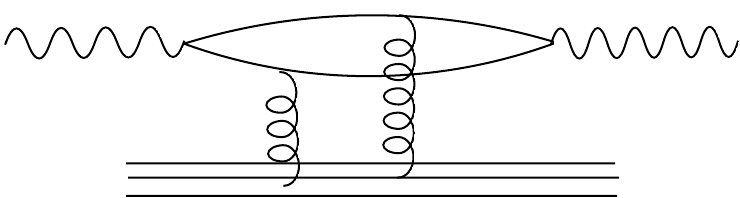}}
\caption{Color dipole picture}
\label{fig:cdp}
\end{figure}

\begin{figure}[h]
{\includegraphics[height=5cm]{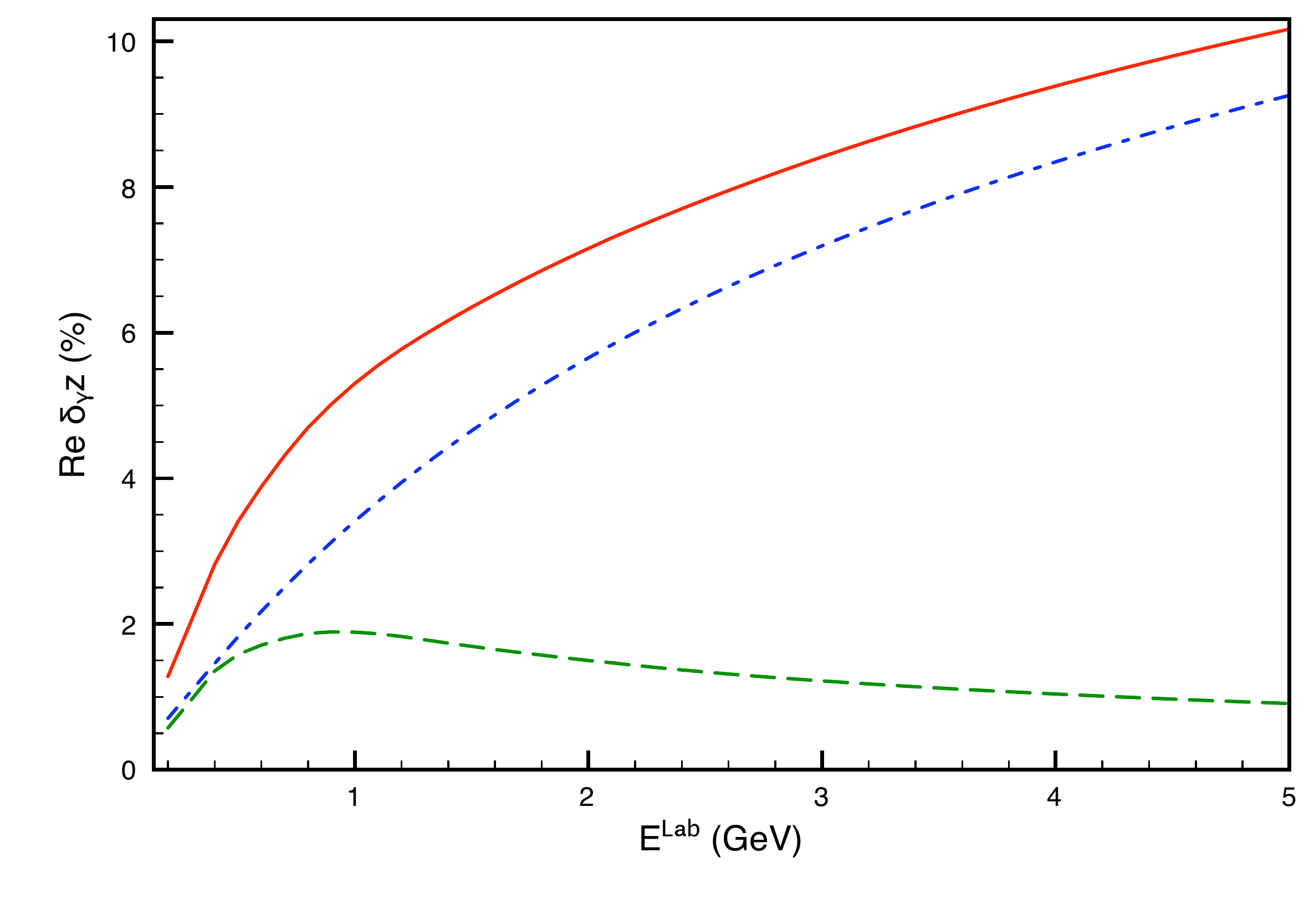}}
\caption{Results for Re$\delta_{\gamma {Z_A}}$ as function of energy. 
The contributions of nucleon resonances (dashed line), Regge (dash-dotted line) 
and the sum of the two (solid line) are shown.
}
\label{fig:delta_gz}
\vspace{-1.2cm}
\end{figure}

Precision tests of the Standard Model at low energies provide an important tool 
to search for New Physics and to constrain model parameters. 
Such tests involve high precision measurements of observables 
that are typically suppressed or precisely vanish in the Standard Model (SM). 
Prominent examples of such observables include the electric dipole moment 
and neutrino magnetic moments.  Another important example of 
a parameter of the nucleon structure suppressed in the SM is the weak charge 
of the proton, $Q_W^p=1-4\sin^2\theta_W$.  With the value of the weak mixing 
angle at low momentum transfers $\sin^2\theta_W(0)=0.23807\pm0.00017$ \cite{erler}, 
the SM predicts the proton weak charge of order $\approx0.05$.  A precise (4\%) measurement of the weak charge of the proton is the aim of the QWEAK experiment at Jefferson Lab \cite{qweak}. \\
\indent
In order to achieve the required precision in the QWEAK experiment, the 
radiative corrections have to be considered.  This was done in various works, 
to mention the most important references \cite{erler,sirlin}, and the 
combined estimate of the theoretical uncertainty is currently 2.2\%. 
This level of precision, coupled with a 2\% measurement of the parity violating asymmetry, would allow for a $0.3\%$ determination of $\theta_W$ at low energies. 
The main difficulty in calculating the radiative corrections originates in 
the hadronic structure-dependent contributions from the box diagrams with the 
exchange of $\gamma\gamma,ZZ,WW$ and $\gamma Z$.
Since the parity conserving amplitude at leading order has a $\frac{1}{Q^2}$ 
pole, the exchange of two photons only leads to a 
correction that vanishes at $Q^2=0$. This amounts in a 
contribution $\sim\alpha Q^2$, with $\alpha\approx\frac{1}{137}$, 
that can safely be neglected. 
Parity violating amplitude in the OBE approximation has no such pole, and 
the respective correction remains finite in the forward direction. 
The $ZZ$ and $WW$-boxes were
estimated in \cite{erler,sirlin} to give a large correction that comes from 
hard exchanged bosons' momenta in the loop, $\sim M_Z(M_W)$, whereas low 
momenta contributions are suppressed by an extra power of $G_F$. 
In this case, all subprocesses 
inside the loop can be treated perturbatively and the contribution can be 
calculated reliably. The situation with $\gamma Z$-box is however more complex, 
since there is generally no reason for hard exchanged momenta to dominate 
the loop with respect to low momenta. 
For atomic PV \cite{sirlin}, it was observed that these dispersion corrections 
are suppressed, as the contributions from the box and the crossed box cancel, 
and the only non-zero term is proportional to the small vector charge of the 
electron thus leading to a correction below 1\%. 
In \cite{erler,mjrm}, this argument was adopted to high energy electrons, 
guided by the assumption of high momentum dominance of the loop integral. 
Clearly, the overall theory uncertainty of 2.2\% relies heavily on this 
cancellation mechanism.
The goal of this letter is to investigate the dispersion correction due to 
$\gamma Z$-box graph in the kinematics of the QWEAK experiment. 
We will provide an explicit calculation of the box and crossed-box corrections in the framework of dispersion relations. \\
\indent
Elastic scattering of massless electrons off a nucleon, 
$e(k)+N(p)\to e(k')+N(p')$, in 
presence of parity violation and in most general form is described by six 
amplitudes $f_i(\nu,t)$, $i=1,2,...,6$, that depend on 
$\nu=\frac{PK}{M}$ with average momenta $K=\frac{k+k'}{2}$ and 
$P=\frac{p+p'}{2}$, and $t=\Delta^2<0$ the elastic momentum transfer,
with $\Delta=k-k'=p'-p$. The basis can be written in the following form:
\beqn
T&=&\frac{e^2}{-t}\bar{u}(k')\gamma_\mu u(k)
\bar{N}(p')\left[f_1\gamma^\mu+f_2i\sigma^{\mu\alpha}
\frac{\Delta_\alpha}{2M}\right]N(p)\nn\\
&+&\frac{e^2}{-t}f_3\,\bar{u}(k')\gamma_\mu\gamma_5 u(k)
\bar{N}(p')\gamma^\mu\gamma_5N(p)\nn\\
&-&\frac{G_F}{2\sqrt{2}}\bar{u}(k')\gamma_\mu\gamma_5 u(k)
\bar{N}(p')\left[f_4\gamma^\mu+f_5i\sigma^{\mu\alpha}
\frac{\Delta_\alpha}{2M}\right]N(p)\nn\\
&-&\frac{G_F}{2\sqrt{2}}f_6\,\bar{u}(k')\gamma_\mu u(k)
\bar{N}(p')\gamma^\mu\gamma_5N(p)\label{eq:ampl},
\eeqn
where only electromagnetic and weak neutral currents are considered. 
The amplitudes $f_{1,2,3}$ are parity conserving (PC), and $f_{4,5,6}$ are 
explicitly parity violating. 
Since we are interested in very forward scattering angles 
$\theta\approx6^\circ$ corresponding to the QWEAK kinematics \cite{qweak}, 
the number of structures that are relevant is further 
reduced. The magnetic terms vanish in the forward direction and can be 
neglected. The Gordon's identity allows to rewrite 
$\bar{N}\gamma^\mu N=\frac{P^\mu}{M}\bar{N}N-\bar{N}i\sigma^{\mu\alpha}
\frac{\Delta_\alpha}{2M}N\to2P^\mu$, where we made use of the nucleon state 
normalization $\bar{N}N=2M$.
The parity conserving amplitude $f_3 $ arises due to an exchange of at 
least two photons or $Z^0$ boson, and is suppressed as 
$\sim{\cal{O}}(t)$, as compared the leading PC amplitude. 
Finally, amplitude $f_6$ depends on nucleon spin and makes no contribution to observables with an unpolarized target.
Only two amplitudes of relevance remain in the forward direction, and denoting 
their forward values as $\tilde{f}_i\equiv f_i(\nu,t=0)$ we obtain
\beqn
T(t\rightarrow 0)=\frac{e^2}{-t}
2\tilde{f}_1\bar{u}(k')\pgdagger u(k)-
\frac{G_F}{\sqrt{2}}\tilde{f}_4\bar{u}(k')\pgdagger\gamma_5u(k)
\label{eq:ampl_forward}
\eeqn
\indent
The form of the forward amplitude coincides with that for spin-0 target 
\cite{ja_chuck}. To leading order in $t$ and in Fermi constant, 
the parity violating asymmetry arises from the interference of PC and PV 
amplitudes,
\beqn
A^{PV}(t\rightarrow 0)=\frac{\sigma_R-\sigma_L}
{\sigma_R+\sigma_L}
=\frac{G_Ft}{4\pi\alpha\sqrt{2}}
\frac{{\rm Re}(\tilde{f}_1^*\tilde{f}_4)}{|\tilde{f}_1|^2}
\eeqn
\indent
At tree level, the amplitudes reduce to $\tilde{f}_1^{OBE}=1$ and 
$\tilde{f}_4^{OBE}=g_A^eQ_W^p=1-4\sin^2\theta_W$, thus the PV asymmetry gives a 
direct access to the weak charge of the proton. Using the exact forward values 
of the amplitudes (the neglect of the $t$-dependence of respective form 
factors) only leads to power corrections in $t$ and can be safely neglected 
\cite{erler}. In the following, we will assume that all the radiative 
corrections except the $\gamma Z$ direct and crossed boxes are known. 
Denoting these other corrections by $\delta_{RC}$, 
we introduce the correction due to $\gamma Z$ exchange as 
\beqn
\delta_{\gamma Z}\equiv
\frac{\tilde{f}_4-\tilde{f}_4^{OBE}}{\tilde{f}_4^{OBE}}-\delta_{RC},
\eeqn
with $\delta_{\gamma Z}$ a complex function energy $\nu$. 
Its real part contributes to the parity violation asymmetry as
\beqn
A^{PV}=\frac{G_Ft}{4\pi\alpha\sqrt{2}}
Q_W^p[1+{\rm Re}\delta_{RC}+{\rm Re}\delta_{\gamma Z}(\nu)]
\eeqn
\indent
To calculate the real part of the $\gamma Z$ direct and crossed box graphs 
shown in Fig. \ref{fig:tbe}
we adopt dispersion relation formalism, and we start with the 
calculation of the imaginary part of the direct box (the crossed 
box contribution to the real part will be calculated using crossing behavior), 
\beqn
{\rm Im}T_{\gamma Z}
\,=\,-\frac{G_F}{\sqrt{2}}\frac{e^2}{(2\pi)^3}\int\frac{d^3\vec{k}_1}{2E_1}
\frac{l_{\mu\nu}\cdot W^{\mu\nu}}{Q^2(1+Q^2/M_Z^2)},
\label{eq:impart}
\eeqn
where $Q^2=-(k-k_1)^2$ 
denotes the virtuality of the exchanged photon and $Z$ (in the 
forward direction they carry exactly the same $Q^2$), and 
we explicitly set the intermediate electron on-shell. In the c.m. of the 
(initial) electron and proton, $E_1=\frac{s-W^2}{2\sqrt{s}}$, with $s$ the 
full c.m. energy squared and $W$ the invariant mass of the intermediate 
hadronic state. Note that for on-shell intermediate states, the exchanged 
bosons are always spacelike. The leptonic tensor is given by
\beqn
l_{\mu\nu}=\bar{u}(k')\gamma_\nu\keldagger_1\gamma_\mu
(g_V^e+g_A^e\gamma_5) u(k).
\eeqn
\indent
In the case of the elastic hadronic intermediate state, the needed structure 
$A(e)\times V(p)$ always contains the explicit factor of $Q_W^p$ or $g_V^e$. 
Correspondingly, the correction $\delta_{\gamma Z}$ is not 
enhanced with respect to the small tree-level coupling, and is generally small, 
in accordance with \cite{erler}. 
We therefore turn to the inelastic contribution. 
In the forward direction, the imaginary part of the 
doubly virtual ``Compton scattering'' ($\gamma^*p\to Z^*p$) amplitude is 
given in terms of the structure functions $\tilde{F}_{1,2,3}(x,Q^2)$, with 
$x=\frac{Q^2}{2Pq}$ the Bjorken variable. Making use 
of gauge invariance of the leptonic tensor, we have
\beqn
W^{\mu\nu}&=&\int d^4xe^{iqx} < p|T\{J_{em}^{\nu\dagger}(x) J_Z^\mu(0)\}|p>\\
&=&2\pi \left\{-g^{\mu\nu}\tilde{F}_1 
+ \frac{P^\mu P^\nu}{Pq}\tilde{F}_2
+ i\epsilon^{\mu\nu\alpha\beta}\frac{P_\alpha q_\beta}{Pq}\tilde{F}_3
\right\}\nn
\eeqn
\indent
Contracting the two tensors, one obtains after little algebra
\beqn
{\rm Im}\delta_{\gamma {Z}}(\nu)&=&\frac{\alpha}{4Q_W^p}
\int_{W^2_\pi}^s\frac{dW^2}{s-M^2}
\int_0^{Q^2_{max}}\frac{dQ^2}{1+\frac{Q^2}{M_Z^2}}\nn\\
&\times&
\left\{g_A^e\frac{1}{Pq}\left[\frac{Pq}{Pk}\tilde{F}_1
+ \left(\frac{2Pk_1}{Q^2}-\frac{P^2}{2Pk}\right)\tilde{F}_2\right]\right.\nn\\
&&\;\;\;-\left.g_V^e\frac{1}{Pq}\frac{(P,k+k_1)}{2Pk}\tilde{F}_3\right\}
\eeqn
with $W^2_\pi=(M+m_\pi)^2$ the pion production threshold, and 
the $Q^2$-integration is constrained below a maximum value
$Q^2_{max}=\frac{(s-M^2)(s-W^2)}{s}$ as a condition of on-shell intermediate 
states for an imaginary part calculation.
In order to write down dispersion relation for the function 
$\delta_{\gamma{Z}}(\nu)$, 
we consider its behavior under crossing. Crossing corresponds to a 
$CP$-transformation applied to a part of the amplitude, so that it relates 
the original reaction $e^-(k)+N(p)\to e^-(k')+N(p')$ to the reaction 
$e^+(-k')+N(p)\to e^+(-k)+N(p')$. The requirement that the crossed 
reaction be described by the same invariant amplitudes taken at the crossed 
kinematics $\nu\to-\nu$ ($K\to-K$ with $P$ unchanged), imposes constraint on 
the form of its $\nu$-dependence. The tensor that multiplies the amplitude 
$\tilde{f}_4$ is even under crossing (being an axial vector). 
Under $C$-parity applied to the electron part, the tree level contribution to 
$\tilde{f}_4$ is also even, and as function of 
$\nu$ the OBE amplitude is an even function - as it is observed, in fact, since 
it only depends on the elastic momentum transfer that is unchanged under 
crossing. 
At one-loop order, exchange of two vector currents (e.-m. and vector part of 
NC) leads to $C$-even behavior, and vector-axial vector exchange to $C$-odd. 
Correspondingly, the part of Im$\delta_{\gamma Z}$ that contains $g_A^e$ is an 
odd function of $\nu$, whereas the one with $g_V^e$ is even.
We distinguish then the two contributions $\delta_{\gamma {Z_V}}$ and 
$\delta_{\gamma {Z_A}}$ that obey dispersion relations of two different forms,
\beqn
{\rm Re}\delta_{\gamma {Z_A}}(\nu)&=&\frac{2\nu}{\pi}\int_{\nu_\pi}^\infty
\frac{d\nu'}{\nu'^2-\nu^2}{\rm Im}\delta_{\gamma{Z_A}}(\nu')\nn\\
{\rm Re}\delta_{\gamma {Z_V}}(\nu)&=&\frac{2}{\pi}\int_{\nu_\pi}^\infty
\frac{\nu'd\nu'}{\nu'^2-\nu^2}{\rm Im}\delta_{\gamma{Z_V}}(\nu')
\label{eq:dr}
\eeqn
\indent
This is just another formulation of the mechanism of the cancellation 
between the box and crossed box graphs observed in \cite{sirlin}. 
For small values of $\nu$ that are relevant for PV in atoms, the explicit 
factor of $\nu$ in front suppresses ${\rm Re}\delta_{\gamma {Z_A}}$. 
Instead, no cancellation occurs for ${\rm Re}\delta_{\gamma {Z_V}}$ which is 
however suppressed by the small value of $g_V^e=-1+4\sin^2\theta_W$. 
The result of Eq. (\ref{eq:dr}) can be considered as a sum rule, since it 
represents the quantity that is to be measured (the proton's weak charge plus 
corrections to it) through other observables (PVDIS structure functions), and 
this relation does not rely on any assumption, other than the neglect of 
higher order radiative corrections. \\
\indent
In absence of any detailed experimental data on PVDIS structure functions, 
to provide estimates of ${\rm Re}\delta_{\gamma {Z_A}}(\nu)$, we need a 
phenomenological model for $W^2$ and $Q^2$ dependence of $\tilde{F}_{1,2}$.
Since we need input from low to high values in both variables, 
one has to distinguish two different contributions: nucleon resonances 
contributions in the intermediate states, and high energy non-resonant part. 
It was shown in a number of papers that a combined color dipole picture/vector 
meson dominance (CDP/VMD) approach allows to describe world data on the 
parity-conserving DIS structure functions $F_{1,2}$ over a wide range. 
Although no data is available to directly judge, whether or not the PVDIS 
structure functions $\tilde{F}_{1,2}$ would follow exactly the same pattern, 
we can expect that they may be similar. 
In CDP, the interaction of the high-energetic photon with the proton target 
occurs in two steps. Firstly, the photon fluctuates into a quark-antiquark pair 
that forms a color dipole. Secondly, this dipole interacts with the target by 
means of exchange of gluons, as shown in Fig. \ref{fig:cdp}.
Correspondingly, the color dipole-proton 
scattering cross section is the only universal non-perturbative ingredient, 
and all the information about the external particles is contained in the 
hadronic ($q\bar{q}$) wave functions of the virtual photon or vector meson, 
depending on the reaction under consideration. In our case, it is the wave 
function of the $Z$-boson, and its vector part is the same as for a virtual 
photon of equal virtuality (which is the case due to forward kinematics), 
except for the quark flavor dependence. This flavor dependence goes as 
$\sum_qe_q^2$ for the $\gamma\gamma$ DIS structure functions, and as 
$\sum_qe_qg_V^q$ for the $\gamma Z$ PVDIS, provided we restrict ourselves to 
the lightest flavors (since all intermediate states have to be real particles, 
production of a heavy quark-antiquark pair requires very high energy) and 
the dependence on quark masses can be safely neglected in the 
photon and $Z$ wave functions. By taking the SM values, 
$\sum_{q=u,d,s}e_q^2=\frac{2}{3}$, whereas 
$\sum_{q=u,d,s}e_qg_V^q=\frac{2}{3}(1+(1-4\sin^2\theta_W))\approx\frac{2}{3}$.
We also remind the reader that the very first PVDIS measurment 
on the deuteron target \cite{pvdis} 
was used to confirm the parton model: in the scaling 
limit, the isoscalar structure functions of DIS and PVDIS depend on the same 
combination of the quark PDFs that cancel out in the asymmetry, leaving only 
kinematical factors and couplings. \\
\indent
This similarity is the case, as well, for the 
$\Delta(1232)$ resonance, as shown in Ref. \cite{sato_lee}. 
Following a simple isospin decomposition, one obtains 
\beqn
<\Delta|J_{NC}^\mu|p>&=&<\Delta|(2-4\sin^2\theta_W)J_{em}^\mu-J_{I=0}^\mu|p>,
\nn\\
\eeqn
with $I$ the strong isospin ($u,d$-quarks are only kept). Since $\Delta$ has 
isospin $3/2$ and the proton $1/2$, 
the isoscalar correction is small, and the NC excitation of the 
$\Delta$ only differs from the electromagnetic one by a factor of 
$2-4\sin^2\theta_W\approx1.08$. \\
\indent
We conclude that the assumption of the 
similarity of the interference $\gamma Z$ structure functions to the usual 
ones with two virtual photons is supported within the parton 
model and CDP at high energy, and at low energies, at least for the most important 
 $\Delta(1232)$ resonance.  We will use this assumption to provide 
a realistic estimate for the dispersion correction to the proton weak charge. Alternatively to the DIS structure functions, one can use the 
transverse and longitudinal virtual photon cross sections $\sigma_{T,L}$ 
related to ${F}_{1,2}$ as (see, e.g., Ref. \cite{tiator} for details)
\beqn
\sigma_T&=&\frac{4\pi^2\alpha}{Pq}F_1,\\
\sigma_L&=&\frac{4\pi^2\alpha}{Pq}
\left[\left(\frac{1}{2x}+\frac{M^2}{Pq}\right)F_2-F_1\right].\nn
\eeqn
\indent
To proceed, we match the relative strength 
of the resonances in Breit-Wigner form and the high energy Regge part to fit 
the real Compton data (for the parameters see \cite{compton_cs}),
\beqn
\sigma_{\gamma p}(W^2,0)&=&\sum_R
\frac{\sigma_R\Gamma_R\Gamma_R^\gamma M_R^2}{(W^2-M_R^2)^2+M_R^2\Gamma_R^2}
+\sigma_{\gamma p}^{Regge}(W^2)\nn\\
\eeqn
\indent
Further, we employ two different forms of $Q^2$ dependence for the two 
contributions. 
Ref. \cite{cvetic} gives analytical form of $W^2$ and $Q^2$ dependence of 
the longitudinal and transverse virtual photon cross sections as
$\sigma_{L,T}^{\it Regge}(W^2,Q^2)\;=\;\sigma_{\gamma p}^{\it Regge}(W^2)
\frac{I_{L,T}(\eta,\eta_0)}{I_T(\eta_0,\eta_0)}$,
with the scaling variable $\eta=\frac{m_0^2+Q^2}{\Lambda^2(W^2)}$ and 
$\eta_0=\eta(Q^2=0)$, so that for the real photons the known Regge asymptotics 
is recovered, and we refer the reader to Ref. \cite{cvetic} for 
further details.
For the resonances, transition form factors are used. The latter 
are to some extent known for a number of resonances and we assume a dipole 
form, $F_T(Q^2)=\frac{1}{(1+Q^2/\Lambda^2)^2}$, and 
$F_L(Q^2)=\frac{Q/\Lambda}{(1+Q^2/\Lambda^2)^{2.5}}$, 
with $\Lambda\approx1$ GeV.
Finally, the form that is used for numerical estimates is 
\beqn
\sigma_{T,L}(W^2,Q^2)&=&\sum_R
\frac{\sigma_R\Gamma_R\Gamma_R^\gamma M_R^2}{(W^2-M_R^2)^2+M_R^2\Gamma_R^2}
F_{T,L}^2(Q^2)\nn\\
&+&\sigma^{Regge}_{T,L}(W^2,Q^2)
\eeqn
\indent
We present results of the dispersion calculation in Fig. \ref{fig:delta_gz}. 
It can be seen that starting from $E_{lab}\approx1$ GeV, high energy (Regge) 
contribution dominates the contribution from the resonances. This is the 
consequence of a relatively slow convergence of the dispersion integral for 
Regge part, while the resonances drop very fast.
In the presented calculation, the upper limit of the integration over $\nu'$ 
was chosen to be 500 GeV, although the $1/\nu'^2$ weighting ensures the 
convergence already at lower values. 
While at very low energies the correction is indeed very 
small, at the 1.16 GeV energy of the QWEAK experiment the correction is 5.7\%. 
More specifically, QWEAK aims at comparing 
the measured weak charge of the proton, 
$\frac{4\pi\alpha\sqrt{2}}{G_Ft}A^{PV}$ to its value as given in SM, 
$Q_W^p\left[1+\delta_{RC}+{\rm Re}\delta_{\gamma Z}\right]$ and from this 
comparison draw conclusions about the New Physics contributions. 
The current estimate of the uncertainty due the the corrections in the square 
brackets is 
2.2\%, and this estimate relies on the assumption that $\delta_{\gamma Z}$ is 
highly suppressed ($\leq$0.65\%). As explained above, this estimate is 
taken over from low energy estimates for PV in atoms, and is not based on any 
microscopic calculation.  Although the numbers presented here are 
themselves model-dependent, our calculation shows that the $\gamma Z$ box 
diagrams can be almost an order of magnitude larger than it was believed to 
date, and this result suggests larger possible theoretical errors for the  
QWEAK experiment. If the uncertainty in the dispersion correction is to be 
comparable to the proposed 2\% experimental error in $A^{PV}$, one may need to 
calculate the dispersion $\gamma Z$ correction (that we think is near 6\%) to a 
fractional accuracy of order 30\%.  Alternatively, uncertainties in these 
dispersion corrections could provide a limit on the precision of a Standard 
Model test.  Since the calculation uses the PVDIS structure functions as input, 
it would be extremely helpful to have experimental data on PVDIS to check the 
model adopted here.
\acknowledgments
This work was supported in part by the US NSF grant PHY 0555232 (M.G.) 
and by DOE grant DE-FG02-87ER40365 (C.J.H.)

\end{document}